# Efficient Indicators to Evaluate the Status of Software Development Effort Estimation inside the Organizations

Elham Khatibi
Department of Information System
Universiti Teknologi Malaysia (UTM)
Skudai 81310, Johor Bahru, Malaysia
`kbelham2@utm.my`

Roliana Ibrahim (Corresponding Author)
Department of Information System
Universiti Teknologi Malaysia (UTM)
Skudai 81310, Johor Bahru, Malaysia
`roliana@utm.my`

*ABSTRACT*

*Development effort is an undeniable part of the project management which considerably influences the success of project. Inaccurate and unreliable estimation of effort can easily lead to the failure of project. Due to the special specifications, accurate estimation of effort in the software projects is a vital management activity that must be carefully done to avoid from the unforeseen results. However numerous effort estimation methods have been proposed in this field, the accuracy of estimates is not satisfying and the attempts continue to improve the performance of estimation methods. Prior researches conducted in this area have focused on numerical and quantitative approaches and there are a few research works that investigate the root problems and issues behind the inaccurate effort estimation of software development effort. In this paper, a framework is proposed to evaluate and investigate the situation of an organization in terms of effort estimation. The proposed framework includes various indicators which cover the critical issues in field of software development effort estimation. Since the capabilities and shortages of organizations for effort estimation are not the same, the proposed indicators can lead to have a systematic approach in which the strengths and weaknesses of organizations in field of effort estimation are discovered.*

*KEYWORDS*

*Software projects, Effort estimation, Framework, Indicator.*

## 1. INTRODUCTION

Project management is one of the most important activities performed throughout the software projects. Main phases of project including analysis, design, implementation and deployment are entirely dependent on project management process. All policies, milestones and responsibilities are organized in project management plan. It is undeniable that planning and scheduling of project is a critical part of project management regardless of project type. In first steps of project, project management team should decide on several important questions related to project planning such as how to arrange development team, how to distribute the responsibilities, how to determine deadline for artifacts, how to determine the duration of project and so on. Appropriate response to these questions can ensure the success of software project. On the other hand, careless answering and lack of attention to planning aspects of project may lead to project fault. Knowledge of project management team regarding the project attributes has a considerable effect on dealing with the mentioned questions.

Development effort is a key attribute of project that influences on most planning and managing aspects. This attribute refers to amount of effort required for project development. It comprises of all activities done within different phases of project. Development effort is basis of decision making on management issues at first steps of project. Accurate forecasting the amount of effort required for performing the project will make the development process so smooth and convenient. This is why so many researchers have tried to increase the accuracy of software development effort prediction using various techniques.

Software projects are strongly different than other projects because the purpose of software projects is producing an intangible product [1-2]. This fact makes the production cycle to be so complicated and difficult in software projects. Therefore, complexity level of software project management is more than other projects. Software project managers are confronted with uncertain and unstable production which is hard to control. Moreover, customer requirements, development technologies and tools are changing rapidly in this field. All of these make the prediction of development effort to be difficult in software projects. As a solution, analyzing of effective factors on development effort estimation can alleviate the problems existing in this area. Investigation of project attributes, limitations, management issues and knowledge of developers in this area can be useful to draw a conclusion in terms of effective factors on management of effort estimation in software projects.

## 2. STUDY BACKGROUND

In 1973, Interactive productivity and Quality (IPQ)[3] was proposed by IBM group as the first automated tool for software development effort prediction. Afterward, Constructive COst Model (COCOMO) was invented by Barry Boehm [4]. COCOMO utilizes some effort drivers to forecast the amount of development effort. It offers several equations based on complexity level of project. "Software Engineering Economics" [4] is a famous book in this area that still numerous researchers employ proposed models in which for effort prediction. Putnam Lifecycle Management (SLIM) [5]and Software Evaluation and Estimation of Resources – Software Estimating Model (SEER-SEM) (Galorath Inc.,1980) have used similar principals to COCOMO [6]. In all mentioned models, Line of Code (LOC) was used for designing the prediction model. In fact, development effort was predicted using LOC as size of project.

Function Point (FP) is so important sizing parameter proposed by Albrecht [7]. It was the first idea for measuring the size of software project by using a functional method. Using of FP showed that it can be placed in effort prediction models instead of LOC because computing process of FP is more reliable and accurate than LOC. Advantages of FP motivated researchers to invent new prediction models based on function point such as Albrecht-Gaffney[7], Kemerer [8] and Matson, Barrett and Mellichamp[9]. Introducing of the new version of COCOMO namely COCOMO II in 2000 [10] is a significant event in this field. COCOMO II considers more details of software project for effort prediction. Prediction equations in this method were improved by applying several scale factors.

In contrary to static methods, there are several dynamic models which rely on using past projects information. Classification And Regression Tree (CART) [11] is one of the dynamic methods in this area. It makes a regression tree according to the available information of completed projects and uses the tree to predict the effort of new project. Analogy Based Estimation (ABE) is the other dynamic method proposed in 1997 [12]. ABE method works based on comparing the attributes of new project and past projects to predict the development effort. It is still so popular because it follows simple and straightforward methods for prediction. ABE have been used widely in recent years [13-16]. Latest advancements in prediction of development effort are related to using of soft computing techniques. Neural networks [13, 17-21] and fuzzy techniques [14, 22-24] are most important soft computing methods employed in this field.

## 3. PRIOR SURVEY-BASED STUDIES

Several studies [25-30] have investigated the accuracy of schedule and effort estimation, which the results showed that 59%-76% of projects exceeded the determined effort and 35%-80% of which exceeded the determined time. Mean value has been utilized in most previous studies to sum up the overruns in time and effort. Exceeding the effort indicated in range of 18% and 41%, while the overrun in time is stated in range of 22% and 25% [25-26, 29-30]. According to the latest Chaos report of Standish group, 32% of software projects are successful, 24% fail and 44% are in challenge.

Since project managers may take small cost and effort overruns [31] easy, it can be helpful to realize the status of effort overruns and recognize the projects which involved in significant effort and cost overruns. Moløkken-Østvold [26] used figures to explain the status of effort overruns which results stated that high number of projects exceeded the determined estimates (below 21%) but only a few projects exceeded effort by higher than 100%. Totally, from this research, it can be said that the mean exceeding in effort (44%) was higher than the median of which (21%). Moløkken-Østvold realized that large projects were more intended to be under estimated. They also investigated if the size of project influenced the accuracy of estimates. It must be said that due to limited size of sample, it was difficult to rely on conclusions from statistical aspect.

Previous surveys [26-27, 32-33] have reported that most projects utilized expert judgment or analogy to estimate the effort while only 14%-26% of which utilized algorithmic estimation techniques. The algorithmic techniques comprise of common models such as, COCOMO, Use-Case models, FP-based models and so on.

Several researches attempted to find the cause for the low acceptance level of algorithmic techniques. For instance, most of algorithmic methods are unable to present enough reliable and accurate estimates [34], many companies do not gather enough data to allow the development of algorithmic models [35], organizations and companies feel not well to utilize techniques that they are unable for fully understand [36] and others.

Prior researches investigated the significance of effort estimation and they achieved approximately the same conclusions. Lederer [28] indicated that almost 84% of the developers ranked effort estimation as "very important" or "moderately important". On the other hand, Moløkken-Østvold [37] indicated that 78% of the respondents rated estimation as "most important", "very important" or "extremely important".

Investigating whether organizations and companies have accepted the existing software effort estimation methods is a critical issue in this field. If they are satisfied, they will have no decision to enhance the estimation methods. Otherwise, they can pay more attention toward its improvement. However, it is not as direct and simple as that. Lederer [28] found that, even though development effort estimation is important, developers are neither specially agree nor disagree with the existing methods. The mean rank was 3.02 on one-to-five point scale (1=very disagree 5=very agree). The author indicated that, in terms of the considerable importance of effort estimation and existing inaccurate estimates, the acceptance of developers indicates that they are satisfied with current methods and they accept the inaccurate estimates.

Moores and Edward [31] indicated that 91% of the responding managers and developers said 'yes' to answer the question 'do you see estimation as a problem?', while only 9% answered 'no'. If this is correct, then, it is true that developers and managers have accepted this problem as a fact of project.

As development phases proceed, the knowledge of developers for software effort estimation becomes more and more, and various estimation techniques are applied at different project stages for any organizations. It is explained in [4] that the uncertainty existing in effort estimates shows a decreasing process as the project proceeds, which is called the Cone of Uncertainty [38] . As an addition, Gryphon stated that the amount of Uncertainty cannot be decreased automatically, but it can be decreased by the accurate estimation techniques as the development phase progresses [39].

This matter is addressed by Lederer [40] and found that 77% projects performed estimation during the primary stages of project, 64% projects performed estimation at feasibility study phase, while 51% within requirements analysis and 48% in requirement design. However, the software project aspects and process have changed significantly since the early 1990's, where the survey by Lederer was conducted.

## **4. PROPOSED FRAMEWORK**

Planning and scheduling of project is a challenging issue for project managers because of uncertain and ambiguous behavior of software projects. The amount of effort is a key factor must be estimated in order to project planning. Since numerous parameters can affect the amount of effort in a software project, classification and prioritization of parameters may facilitate the effort estimation process. Managers need to know the importance of each parameter to make decision realistically throughout the project planning. Each parameter is related to a part of software project, and it influences on a part of activities, artifacts and roles.

Proposing a framework needs to determine the exact scope and area which must be investigated through the survey. In this research, we are going to focus on some aspects of software projects that may affect the effort estimation (based on the results obtained from the prior studies). As seen in Figure 1, knowledge of developers in terms of effort estimation, limitation and obstacles against accurate estimation, importance level of project attributes as well as management issues are the main issues that must be assessed inside the organizations to clarify the situation of organization in terms of effort estimation. In the following section, some indicators are proposed to assess the different parts of the mentioned framework.

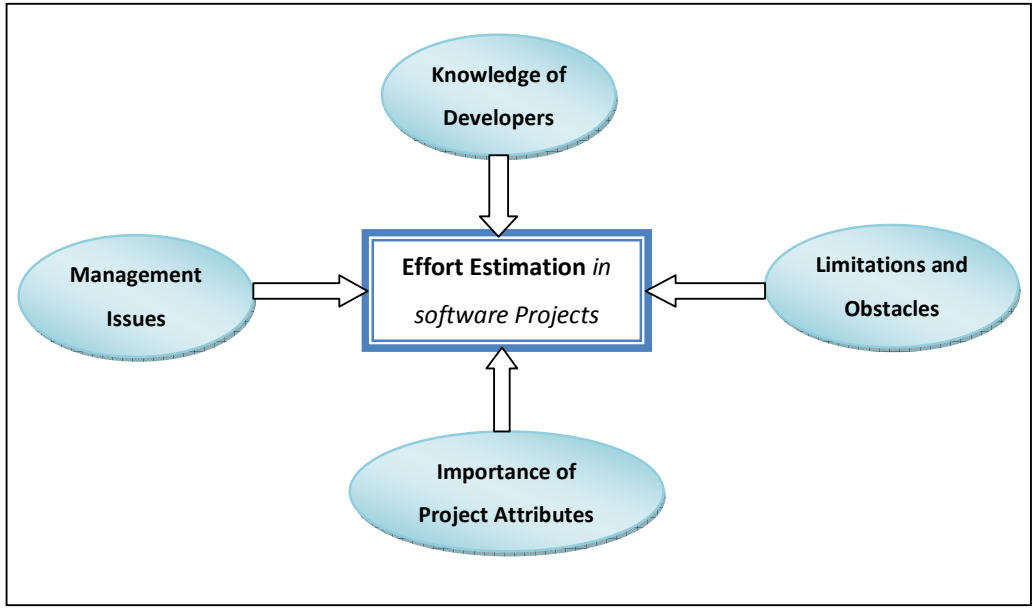

Figure 1. The investigation framework

For the issues mentioned as the important parts of effort estimation inside the organizations, the measurement procedure must be explained to ensure the applicability of the method. The indicators are utilized to assess and investigate the related case. In order to find the most suitable indicators several critical questions are considered. For example, how the survey wants to examine the knowledge of developers in field of effort estimation? Which limitations and obstacles are considered in the survey? Which project attributes are involved in this research? and so on. The indicators are determined so that the investigation results can answer the questions. Figure 2 displays the indicators we have determined to evaluate the different parts of effort estimation.

### 4.1 Knowledge of Developers

Regarding the knowledge of developers, it is very important for managers to know how developers are familiar with the different aspects of effort estimation. This can be known by investigating the knowledge of developers in terms of the process of effort estimation. In addition, the familiarity level of developers with the latest effort estimation methods is a critical issue to examine the capability of developer for effort estimating. Finally, the prior experience of developers is an undeniable factor determination of developer's capability in this field.

### 4.2. Management Issues

Regarding the management issues, it must be evaluated that how managers believe to effort estimation. If they do not believe the estimation, they may force the team to determine the effort less than the most likely effort. Managers must be aware of the benefits of accurate effort estimation.

Attention to effort estimation through the management activities must be evaluated inside the organizations. Some indictors such as clearly define activities for effort estimation, allocate staff to conduct the effort estimation, define milestone and plan for effort estimation and  continuous training of developers in field of latest effort estimation tools and methods must be considered here. Creating a database of historical project effort factors and documenting the process of effort estimation are the other factors in this field. Team organization and coordination as the other important indicators are considered in the proposed framework. Analysis and determine the possible factors lead to inaccurate estimates is the other indicator must be considered by managers. The mentioned factors can be unstable demand, change the development process, lack of historical project information as a basis for estimation, the lack of monitoring of the effort. The last indicator in this group is monitoring. Timely adjustments must be performed to estimate the target. According to the software project's progress, the estimated effort must be adjusted to achieve the required accuracy. Effort estimates must be evaluated by an independent person. In addition, effort estimates must be accurately recorded and the change of accuracy and improvement must be continuously controlled.

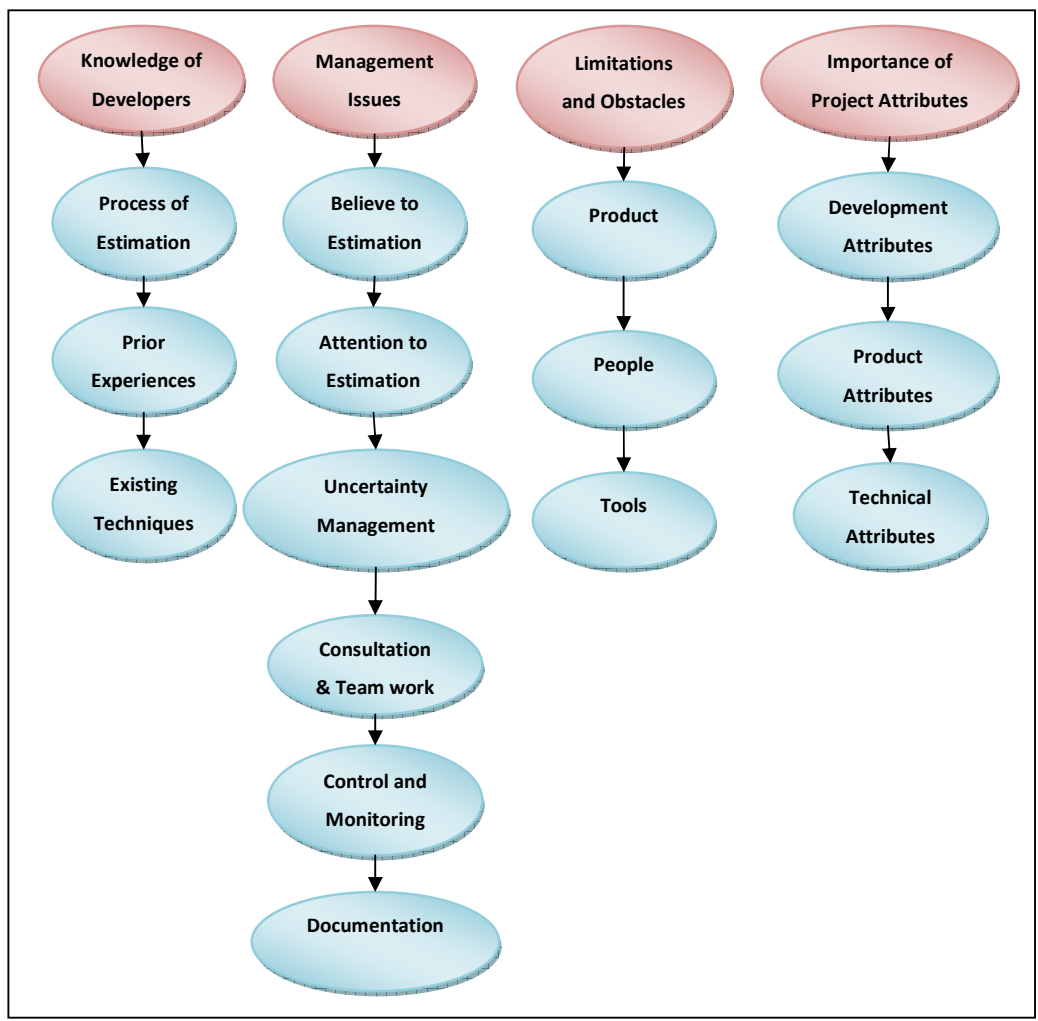

Figure 2. The proposed indicators

**4.3 Limitations and Obstacles**

As stated in the previous sections, effort prediction is a challenging and complicated process in the software projects. There are some factors and reasons which make the effort prediction to be very difficult. This group of indicators includes some of the most important factors and obstacles lead to inaccurate estimates inside the organizations. These indicators have been divided into three main groups: product, people and tools. Frequent changes in a software requirements, unclear and vague software requirements and lack of historical project data are the obstacles related to the product group. Lack of appropriate estimation methods or estimation process, lack of the support of the estimation tools and lack of required information which must be used by tools are obstacles related to the group of tools. Not enough time or manpower to carry out the effort estimate, pressure from senior managers, customers, or others, directly specify or modify the estimation results, lack of participation of application developers, lack of timely supervision and control of cost according to plan, lack of analysis of software systems and the associated risks, lack of coordination among the relevant stakeholders of the customers, users, system design and development and lack of risk analysis and management of software projects are the obstacles related to the people group.

**4.4 Importance of Attributes**

There are several standard and defined attributes for any software projects, which include organization type, development type, development technique, development style, application type, programming language, CASE tools as well as size. These attributes need to be investigated in order to clarify that how they influence on project effort. In order to discover the effect of these attributes on project effort, a comprehensive analysis must be performed inside the organization. Various types of the software projects and the large number of attributes make the analysis to be complicated and time consuming. In order to overcome the complexity of this problem, we have classified the related attributes into three main groups of development, product and technical. Selection of attributes has been performed based on the importance and worth of each attribute in terms of project effort. Prior studies and interview are the main instruments helped us to select the attributes.

## 5. CONCLUSION

Software is an important concept in the modern business, government and military operations. This indicates that hundreds of new applications are produced and hundreds of existing applications are modified every year either by a corporation or a state government. Huge host of software projects in the today's business world means that software effort estimating is now a significant activity for any company that produces or develops software. Combined with software development process, software effort estimation process can help projects to provide credible and reliable plans to develop the software requirements and satisfy agreements. It can also help other project activities particularly management issues, by presenting accurate and timely effort estimates throughout the project. Lack of the analytical and survey-based studies is the problem behind the inaccurate estimation of software development effort. The numeral and quantitative estimation methods cannot overcome the non-normality of software projects because the accuracy of estimates strongly depends on the management issues which must be evaluated and improved inside the organizations. The management issues are different from one organization to another one and a unified evaluation framework can be a suitable solution to this problem. This paper proposed a framework including several indicators to evaluate the real situation of effort estimation process in organizations. The indicators were classified into four main groups so that they covered the most important issues related to the effort. The measurement procedure for the indicators located in four groups was explained separately to ensure the ability of the framework to be implemented. This framework can be helpful for managers to know the strengths and weaknesses of organization regarding the process of effort estimation. On the other hand it can be suitable to find a unified method to evaluate and improve the status of effort estimation in different organizations. Conduction of a survey using the framework proposed in this study is the future work we are going to do.

## ACKNOWLEDGEMENTS

Special thanks to the Universiti Teknologi Malaysia for financing and funding this research through Research University Grant.

## REFERENCES

[1]   G. Stepanek, Software Project Secrets: Why Software Projects Fail USA: Apress, 2005.

[2]   V. Khatibi.B and D. N. A. Jawawi, Software Cost Estimation Methods: A Review, Journal of Emerging Trends in Computing and Information Sciences, vol. 2, pp. 21-29, 2011.

[3]   C. Jones, Estimating software costs: Bringing realism to estimating, 2nd ed. New York: NY: McGraw-Hill, 2007.

[4]   B. W. Boehm, Software engineering economics. Englewood Cliffs: NJ: Prentice Hall, 1981.

[5]   L. H. Putnam, A General Empirical Solution to the Macro Software Sizing and Estimating Problem, IEEE Transactions on Software Engineering,, vol. SE-4, pp. 345-361, 1978.

[6]   B. W. Boehm and R. Valerdi, Achievements and Challenges in Cocomo-Based Software Resource Estimation, IEEE Softw., vol. 25, pp. 74-83, 2008.

[7]   A. J. Albrecht and J. A. Gaffney, Software function, source lines of codes, and development effort prediction: a software science validation, IEEE Trans Software Eng. SE, vol. 9, pp. 639-648, 1983.

[8]   C. F. Kemerer, An empirical validation of software cost estimation models, Commun. ACM, vol. 30, pp. 416-429, 1987.

[9]   J. E. Matson, et al., Software Development Cost Estimation Using Function Points, IEEE Trans. Softw. Eng., vol. 20, pp. 275-287, 1994.

[10]  B. Boehm, Software Cost Estimation With COCOMO II: Prentice Hall, 2000.

[11]  L. Breiman, et al., Classification and Regression Trees: Pacific Grove, CA: Wadsworth, 1984.

[12]  M. Shepperd and C. Schofield, Estimating Software Project Effort Using Analogies, IEEE Transaction on software engineering,, vol. 23, pp. 736-743, 1997.

[13]  Y. F. Li, et al., A study of the non-linear adjustment for analogy based software cost estimation, Empir Software Eng, vol. 14, pp. 603-643, 2009.

[14]  M. Azzeh, et al., Fuzzy grey relational analysis for software effort estimation, Empirical Software Engineering, vol. 15, pp. 60-90, 2010.

[15]  Q. Song and M. Shepperd, Predicting software project effort: A grey relational analysis based method, Expert Systems with Applications, vol. 38, pp. 7302-7316, 2011.

International Journal of Managing Information Technology (IJMIT) Vol.4, No.3, August 2012

[16] C.-J. Hsu and C.-Y. Huang, Comparison of weighted grey relational analysis for software effort estimation, Software Quality Journal, vol. 19, pp. 165-200, 2011.

[17] C. S. Reddy and K. Raju, A Concise Neural Network Model for Estimating Software Effort, International Journal of Recent Trends in Engineering, vol. 1, pp. 188-193, 2009.

[18] I. Kalichanin-Balich and C. Lopez-Martin, Applying a Feedforward Neural Network for Predicting Software Development Effort of Short-Scale Projects, presented at the Software Engineering Research, Management and Applications, 2010.

[19] J. Kaur, et al., Neural Network-A Novel Technique for Software Effort Estimation, International Journal of Computer Theory and Engineering, vol. 2, pp. 17-19, 2010.

[20] I. Attarzadeh and O. S. Hock, Proposing a new software cost estimation model based on artificial neural networks presented at the Computer Engineering and Technology (ICCET),2nd International Conference on 2010.

[21] R. Bhatnagar, et al., Software Development Effort Estimation Neural Network Vs. Regression Modeling Approach, International Journal of Engineering Science and Technology, vol. 2, pp. 2950-2956, 2010.

[22] P. R. P.V.G.D, et al., Fuzzy Based Approach for Predicting Software Development Effort, Software Engineering(IJSE), vol. 1, pp. 1-11, 2010.

[23] K. R. Ch. Satyananda Reddy, Improving the accuracy of effort estimation through Fuzzy set combination of size and cost drivers, WSEAS Transactions on Computers,, vol. 8, pp. 926-936, 2009.

[24] H. K. V. Vishal Sharma, Optimized Fuzzy Logic Based Framework for Effort Estimation in Software Development, International Journal of Computer Science, vol. 7, pp. 30-38, march 2010 2010.

[25] A. M. Jenkins, et al., Empirical Investigation of Systems Development Practices and Results, Information & Management, vol. 7, pp. 73-82, 1984.

[26] K. Moløkken, et al., A survey on software estimation in the Norwegian industry, presented at the 10th International Symposium on Software Metrics, 2004.

[27] F. J. Heemstra, Software cost estimation, Information and Software Technology, vol. 34, pp. 627-639, 1992.

[28] A. L. Lederer and J. Prasad, Causes of Inaccurate Software Development Cost Estimates, Journal of Systems and Software, vol. 31, pp. 125-134, 1995.

[29] F. Bergeron and J.-Y. St-Arnaud, Estimation of Information Systems Development Efforts: A Pilot Study, Information & Management, vol. 22, pp. 239-254, 1992.

[30] C. Sauer and C. Cuthbertson, The State of IT Project Management in the UK 2002-2003, Templeton College, University of Oxford, 2003.

[31] T. T. Moores and J. S. Edwards, Could Large UK Corporations and Computing Companies Use Software Cost Estimating Tools?  A Survey, European Journal of Information Systems, vol. 1, pp. 311-319, 1992.

[32] G. Wydenbach and J. Paynter, a Survey of Practices in New Zealand Software Project Estimation, Zealand Journal of Computing, vol. 6, pp. 317-327, 1995.

[33] K. McAulay, Information Systems Development and the Changing Role of MIS in the Organisation, presented at the First New Zealand MIS Management Conference, Wellington, 1987.

[34] L. C. Briand , et al., COBRA: A hybrid method for software cost estimation, benchmarking and risk assessment, presented at the Proceedings of 20th International Conference on Software Engineering, 1998.

[35] M. Ruhe, et al., Cost estimation for web applications, presented at the 25th International Conference on Software Engineering, Portland, Oregon, 2003.

[36] M. Jorgensen, A review of studies on expert estimation of software development effort, Systems and Software, vol. 70, pp. 37-60, 2004.

[37] K. Molokken and M. Jorgensen, A review of software surveys on software effort estimation, in Empirical Software Engineering, Washington, DC, USA, pp. 223-230, 2003.

[38] S. McConnell, Rapid Development: Taming Wild Software Schedules. : Microsoft Press, 1996.

[39] S. Gryphon, et al., Letters: The Cone of Uncertainty, IEEE Software, vol. 23, 2006.

[40] A. L. Lederer and J. Prasad, Nine Management Guidelines for Better Cost Estimating, Communications of the ACM., vol. 35, pp. 51-59, 1992.